\documentclass[journal, twocolumn]{IEEEtran}
\usepackage{amsfonts}
\usepackage{amssymb}
\usepackage{graphicx}
\usepackage{amsmath,amssymb,amsthm} 
\usepackage{mathtools}
\usepackage{amsmath, amsfonts,epsfig, multirow, floatflt}
\usepackage{epsfig,graphics,graphicx,latexsym,amsfonts,amssymb,amsmath,verbatim}
\usepackage{subfigure}
\usepackage{cite}
\usepackage{color}
\usepackage{soul}
\usepackage{multirow}
\usepackage{array}
\usepackage{algorithm}
\usepackage{algpseudocode}
\usepackage{setspace}
\usepackage{booktabs}
\usepackage{threeparttable}
\usepackage{makecell}
\usepackage{epstopdf}
\usepackage{tabularx}
\usepackage{mathrsfs}
\usepackage{multirow}

\usepackage{amsmath}
\usepackage{algorithm}
\usepackage{algorithmicx}

\usepackage{amsmath}
\usepackage{txfonts}
\graphicspath{{figure/}}

\begin{document}
	
\title{Futures-Based Resource Trading and Fair Pricing in Real-Time IoT Networks}

\author{Shuran~Sheng,  Ruitao~Chen, Peng~Chen,~\IEEEmembership{Member,~IEEE,}           Xianbin~Wang,~\IEEEmembership{Fellow,~IEEE}, and Lenan~Wu
	\thanks{S. Sheng, R. Chen and X. Wang are with Department of Electrical and Computer Engineering, Western University, London, ON N6A 5B9, Canada (e-mail: {ssheng23, rchen328, xianbin.wang}@uwo.ca).}
	\thanks{S. Sheng and L. Wu are with the School of Information Science and Engineering, Southeast University, Nanjing 210096, China (e-mail: shengshuran@seu.edu.cn; wuln@seu.edu.cn).}
	\thanks{P. Chen is with the State Key Laboratory of Millimeter Waves, Southeast University, Nanjing 210096, China (e-mail: chenpengseu@seu.edu.cn).}
	\thanks{Corresponding author: Xianbin~Wang.}}
\maketitle

\begin{abstract}

We propose a futures-based resource trading scheme via a forward contract to tackle the risk of trading failure and unfairness associated with the on-site negotiation process in facilitating resource sharing in wireless networks. More specifically, the resource requester and the resource owner negotiate a mutually beneficial forward contract in advance, where the agreement between the two parties are based on the historical statistics related to the resource supply and demand. The risk of trading failure is utilized to determine the contract price and resource amount. Spectrum trading between two different service providers is studied as an example and simulation results show that the proposed futures-based resource trading scheme achieves better performance in terms of success rate and fairness compared with the traditional on-site mechanism. \par
\end{abstract}


\section{Introduction}
\label{SEC_INTR}

\IEEEPARstart{T}{echnology} advancements on sensing, communications, and computing directly accelerate the recent development of Internet of Things (IoT), leading to rich and diverse applications
in industry and business processes~\cite{A_Al_Fuqaha}. For enabling IoT applications, smart devices and networks rely on the capabilities for information collection, processing and tight collaboration with their neighbors. However, the rapid proliferation of IoT devices and real-time IoT applications result in dramatically increased demand for communication and computing resources, posing a significant challenge for the underlying resource constrained wireless networks. Due to their resource constraint, it is extremely difficult for such devices and networks to meet stringent requirements of many real-time IoT applications. To overcome these challenges, resource sharing among IoT devices and networks has been studied as an effective solution by exploiting unused resources~\cite{N_Correia}.\par

However, one major difficulty of the resource sharing is how to design an incentive mechanism by encouraging the selfish resource owners to fully share their under-utilized resources with others. As a result, resource trading has been studied to enable resource selling and buying among resource owners and resource requesters.\par

To enable resource trading, price and amount for resource sharing need to be negotiated among different trading entities to maximize the revenue of the resource owners while meeting the demands of requesters. Most studies in this area utilize on-site trading~\cite{Y_Cao,B_Xia,I_Bajaj}, where negotiations, including the trading price and resource amount,  are based on the dynamic available resource and changing network conditions. All requesters compete with each other for the shared resources in the on-site trading market via game theory (e.g., auction and Stackelberg game)~\cite{Y_Cao,I_Bajaj}. As a result, some requesters have the risk of failure to access  the resources, leading to the violation of QoS.  Additionally, with the random nature of available resource and demand,  on-site resource trading often features drastically fluctuating price and inevitable unfairness due to stringent trading latency requirement in the real-time IoT applications. 
\par

In order to reduce the probability of trading failure while improving the trading fairness, \emph{\textbf{futures}}-based resource trading scheme in real-time IoT networks is proposed in this letter. \emph{\textbf{Futures}} refers to a forward contract that different entities agree to buy or sell a product at a specific price in the future~\cite{JC_Hull}, and has been widely adopted in financial and commodity exchange to reduce the risk of dynamic price and resource availability ahead of the actual transaction~\cite{Tsay, SE_Khatib}. Different from the existing on-site resource trading, the proposed new futures-based resource trading scheme regulates the price and quantity prior to the on-site request through a mutually beneficial forward contract, while considering the uncertainty of resource supply and demand. Specifically, the utility of the resource owner and the resource requester are formulated according to the dynamic network conditions. The price and amount of the futures contract are set based on the risk estimation for trading failure, which is incurred by the dynamic resource availability and network conditions. Spectrum trading between two different service providers is studied to evaluate the proposed futures-based trading mechanism.  \par

\section{Futures-Based Resource Trading Scheme}
\label{ }
Futures-based resource trading scheme is first presented in this Section. Spectrum trading is then studied as an example to illustrate the risk analysis and contract establishment.\par
\subsection{Futures-Based Resource Trading Scheme}
Resource trading is an effectively approach to exploit underutilized resources to meet the requirements of real-time IoT applications in resource constrained networks. However, the existing resource trading scheme determines price and amount for resource trading via on-site auction and game-theoretic approach. In that case, some urgent requesters have high risk of unable to obtain the resources on time. Furthermore, On-site trading could also bring trading unfairness due to imbalanced resource availability and demand. Therefore, futures-based resource trading is proposed to address these issues, in which the resource requester and the resource owner negotiate a mutually beneficial forward contract on the trading price and amount in advance. The utilities of the two trading sides are random because of the changing resource availability and demand. In this letter, the trading statistics is assumed to be known based on the historical records. The risk of trading failure is analyzed to determine the fair resource price and amount of the futures-based contract. In IoT networks, the risk may come from the randomness of resource availability, dynamic demand, varying channel quality, and so on. The contract requester will access the shared resource based on the mutually beneficial agreement.\par
\subsection{Resource Trading Model for Spectrum Sharing}
\begin{figure}[htb!]
	\setlength{\abovecaptionskip}{-0cm}
	\setlength{\belowcaptionskip}{-0cm}
	\centering
	\includegraphics[width=0.48\textwidth]{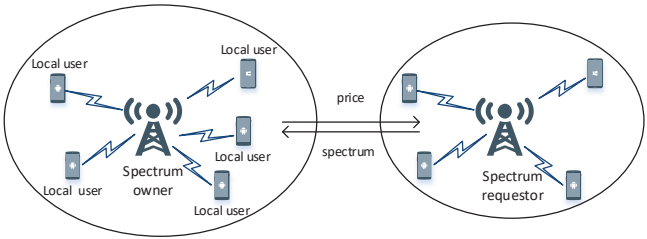}  
	\caption{Futures-based spectrum trading between two service providers.}
	\label{system_model}
\end{figure}

The ever increasing density of IoT devices has posed growing demand for spectrum resources. Futures-based spectrum resource trading is studied here between two different service providers, as shown in Fig.~\ref{system_model}. Each service provider is corresponded to an access point (AP), such as a base station and a Wi-Fi access point. With total bandwidth of $W$, the under-utilized spectrum of the service provider (resource owner) could sell its idle spectrum to the over-utilized service provider (resource requester). The two sides determine the price $p$ and the amount $r_s$ for trading via a mutually forward contract for future spectrum sharing.\par

\subsection{Utility and Risk of the Spectrum Owner}
The utility of the spectrum owner is the profit that can be obtained, including the revenue from local user services, the revenue from spectrum selling to the spectrum requester, and the cost incurred by the performance degradation due to spectrum sharing~\cite{D_Niyato}.\par
\begin{equation}\label{eq_1}
	\begin{aligned}
	U_b(p,r_s,n_c) =  c_1 n_c+pr_s -c_2 n_c\left(B_{req}-k_c\frac{W-r_s}{n_c}\right),
	\end{aligned}
\end{equation}
\noindent where $n_c$ is the number of local users served by spectrum owners that influences the amount of available spectrum for trading. Based on historical statistics, the value of $n_c$ can be obtained. $c1$ and $c2$ are weights for the revenue from local users and cost due to the performance degradation, respectively.  $B_{req}$ and  $k_c$ denote the required bandwidth and spectral efficiency of wireless transmission for local user services. The remaining spectrum for the local services is reduced due to the resource sharing, resulting in performance degradation of the network. \par

The risk of the spectrum owner comes from the prediction uncertainty on the number of local user number $n_c$, which is defined as the probability that the ratio of the random utility and its mean value $E\left[U_b(p,r_s,n_c)\right]$ is less than a threshold $\rho_b$.
\begin{equation}\label{eq_2}
	\begin{aligned}
	R_b(p,r_b,n_c) =  prob \left\{\frac{U_b(p,r_s,n_c)}{E[U_b(p,r_s,n_c)]}\le \rho_b \right\}.
	\end{aligned}
\end{equation}
The spectrum owner provides an acceptable tolerance for the risk to reach the contract. Therefore, the objective of the spectrum owner is to maximize its expected utility with the constraint of the risk threshold of trading failure $T_b$, which is
\begin{equation}\label{eq_3}
	\begin{aligned}
	\max_{p,r_s} \quad & E\left[U_b(p,r_s,n_c)\right]   \\
	\mbox{s.t.}\quad \, \, \, \,
	& R_b(p,r_b,n_c) \le T_b.
	\end{aligned}
\end{equation}
\subsection{Utility and Risk of the Spectrum Requester}
The utility of the spectrum requester includes spectrum revenue and the payment for accessing the spectrum, which is affected by the wireless transmission. The wireless transmission rate can be dynamically adjusted by adaptive modulation based on the channel quality. Therefore, the utility of the spectrum requester is expressed as~\cite{D_Niyato}
\begin{equation}\label{eq_4}
	U_d(p,r_s,\gamma)=\omega_n \log_2(1+k_dr_s) -pr_s,
\end{equation}
\noindent in which $\omega$ is the revenue weight, $k_d$ is the spectral efficiency. The value of $k_d$ of the requester can be given as~\cite{D_Niyato}.
\begin{equation}\label{eq_5}
	\begin{aligned}
	k_d=\log_2(1+ K \gamma),K=\frac{1.5}{\ln(0.2/BER^{tar})},
	\end{aligned}
\end{equation}\par

\noindent where $\gamma$ is the signal-to-noise ratio (SNR), which is a random variable, and $BER^{tar}$ is the target bit-error-rate (BER).\par

When the SNR is too small, the utility that the requester could earn may be negative and thus the contract cannot be reached. Therefore, the minimum utility is set to zero to guarantee the minimum QoS requirement.
\begin{equation}\label{eq_6}
	U_d^{min}(p,r_b,\gamma)=0.
\end{equation}

The risk of the spectrum requester is incurred by the prediction uncertainty of the randomness of the value of $\gamma$. So it can be defined as the probability that the utility will be too close to its minimum
\begin{equation}\label{eq_7}
	\begin{aligned}
	R_d(p,r_s,\gamma)=prob \left\{\frac{U_d(p,r_s,n_c)}{U_d^{min}(p,r_b,\gamma)}\le 1+\rho_d\right\}.
	\end{aligned}
	\end{equation}
Similarly, the spectrum requester also has a risk tolerance to accept the contract. Therefore, the objective of the spectrum requester is to maximize its expected utility with the constraint of the risk threshold of trading failure $T_d$, and is given as
	\begin{equation}\label{eq_8}
	\begin{aligned}
	\max_{p,r_s} \quad & E	\left[ U_d(p,r_s,\gamma) \right]   \\
	\mbox{s.t.}\quad \, \, \, \,
	& R_d(p,r_s,\gamma) \le T_d.
	\end{aligned}
	\end{equation}
\section{Contract Negotiation of the Price and Resource Amount for Futures Based Trading}
\label{ }
The price and amount for resource trading are negotiated within the contract period, which can be predicted according to the resource usage of the owner and requester. Noted that the problem of the contract period is out of the scope of this letter. The two sides only negotiate for the price and resource amount of trading, which is performed in an iterative process, as shown in Algorithm \ref{alg:pricing}. In each iteration, the spectrum owner proposes a contract price $p$ and computes corresponding acceptable amount range for trading based on its risk estimation. Upon receiving the asking price, the spectrum requester performs its risk analysis to identify an affordable range of trading amount. If the two ranges overlap, the requester accepts the price at the amount that maximize its utility. After all acceptable price and amount pairs are found, the spectrum owner sets the final contract price and amount pair for trading that maximizes its utility. If there is no overlap after sufficient rounds of iterations, the negotiation terminates, which means the trading fails.\par

\begin{algorithm}[!htb]
	\caption{Futures-based Pricing Algorithm}
	\label{alg:pricing}
	\begin{algorithmic}[1]
	    \State \textbf{Require}: The spectrum owner sets the minimum spectrum price $p_{min}$ and its risk of trading failure threshold $T_b$, the spectrum requester sets its risk of trading failure threshold $T_d$ .
		\State \textbf{Repeat}:
		\State The spectrum owner announces its price $p$ and corresponding range of trading amount $r_b$  under $T_b$ accroding to \eqref{eq_2}.
		\State The requester estimates its risk based on \eqref {eq_7} and finds its acceptable spectrum size range of $r_d$ under $T_d$. \label{marker1}
		\If {$r_b$ and $r_d$ overlap}
		\State The requester finds the spectrum amount $r$ that maximizes its expected utility.
		\EndIf	
		\State The owner updates the price $p=p+\Delta p$.
		\State \textbf{Until}: $r_b$ and $r_d$ do not overlap.	
		\State The owner finds the optimal amount $r_{op}$ price $p_{op}$ among all $r$ that maximizes its expected utility from all acceptable price and quantity of the requester.
		\State \textbf{Return}: $r_{op}$ and $p_{op}$.
	\end{algorithmic}
\end{algorithm}
\section{Simulation Results and Analysis}
\label{ }
In the simulation, we set the total spectrum bandwidth $W$ as 30MHz, the number of local users follows Poisson distribution with average being 8. The required spectrum bandwidth of each local user $B_{req}$ is assumed as 1 MHz, the SNR follows uniform distribution between 9 dB to 22 dB. Furthermore, the risk threshold of the spectrum owner is set to 0.2. \par

\begin{figure}[htb!]
	\setlength{\abovecaptionskip}{-0cm}
	\setlength{\belowcaptionskip}{-0cm}
	\centering
	\includegraphics[width=0.50\textwidth]{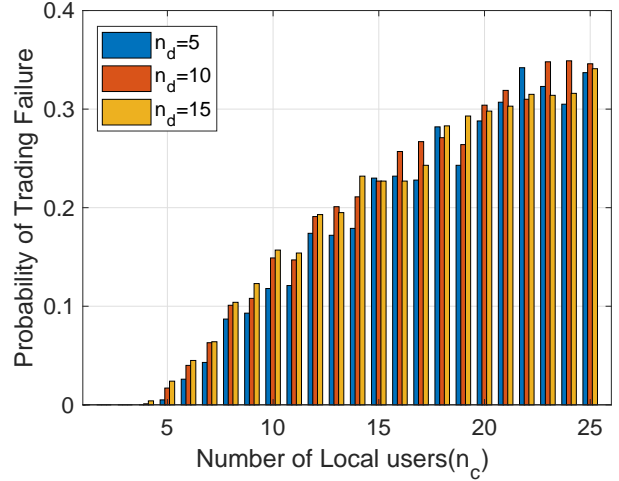}  
	\caption{Probability of trading failure comparison between the proposed futures-based trading scheme and on-site trading mechanism.}
	\label{prob_of_trade_failure}
\end{figure}
A contract user obtains the spectrum according to the forward contract, which indicates that there is no risk of trading failure in the futures-based trading. However, the on-site requester has risk of failure to access the resource due to the competition with others. Therefore, the probability of trading failure in the on-site trading is the same with the performance improvement in the futures-based trading scheme. Fig.~\ref{prob_of_trade_failure} shows the probability of trading failure in the on-site trading scheme ~\cite{N_Sawyer}, where the probability of trading failure is 0 at the beginning and then steps up with increasing number of local users. This can be mainly attributed to the fact that the available spectrum is sufficient for trading when there are a few local connections. However, the available spectrum cannot meet the QoS requirement of the on-site requester compared to the contract user when the number of local users continues to increase.\par
{\color{red} 
\begin{figure}[htb!]
	\setlength{\abovecaptionskip}{-0cm}
	\setlength{\belowcaptionskip}{-0cm}
	\centering
	\includegraphics[width=0.50\textwidth]{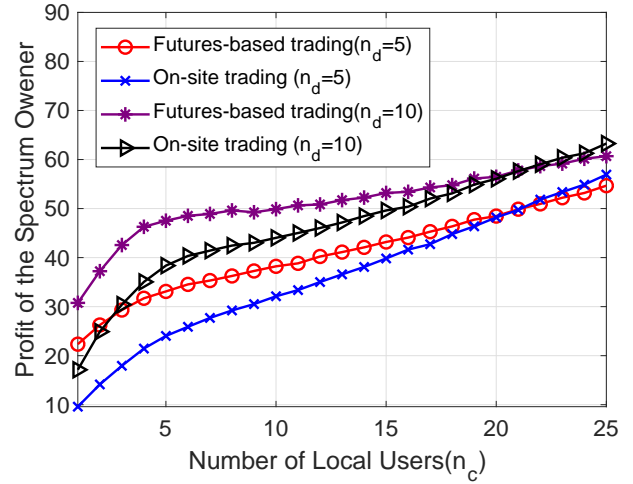}  
	\caption{Comparison of profit of the owner between the on-site and the proposed futures-based trading mechanism.}
	\label{owner_profit}
\end{figure}}
As shown in Fig.~\ref{owner_profit}, the futures-based trading achieves higher profit compared to that in the on-site trading in most cases. The profit increases faster at the beginning, while the gap between the two trading schemes becomes smaller. This indicates that the punishment is small when there are a small number of local users. The available spectrum for trading is decreased with increased local users, leading to slower growth for spectrum utilization. In contrast, the futures-based trading scheme shows worse performance than the on-site trading mechanism as the number of local users continues to increase.\par 

The trading fairness can be reflected by the stability of the spectrum revenue, which is related to the allocated spectrum size and spectrum efficiency. We define that the trading fairness is reciprocal of the variance of the spectrum revenue.\par
	\begin{equation}
	F=\frac{1}{var(\log_2(1+k_dr_s))}.
	\end{equation}
\begin{figure}[htb!]
	\setlength{\abovecaptionskip}{-0cm}
	\setlength{\belowcaptionskip}{-0cm}
	\centering
	\includegraphics[width=0.50\textwidth]{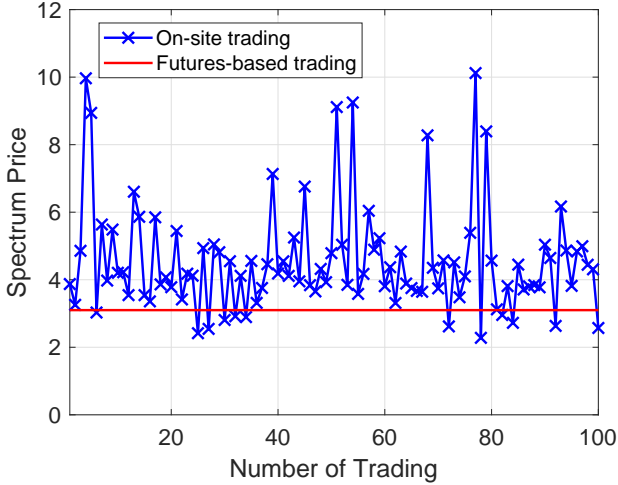}  
	\caption{Spectrum price in the on-site and futures-based trading.}
	\label{Spectrum_Price}
\end{figure}
From the above equation, we can observe that the larger fluctuation of the amount for trading will introduce the worse fairness. In fact, the fluctuation of the spectrum quantity is affected by the variation of the price. Fig.~\ref{Spectrum_Price} shows the spectrum price of the futures-based and the on-site spectrum trading. It can be seen futures-based spectrum price remains constant, whereas the price in on-site trading fluctuates significantly. Furthermore, spectrum price in the futures-based trading is almost lower than that in the on-site trading. The reason is that the generally insufficient spectrum results in higher on-site price in most cases compared to the contract price which is determined by statistics.  In addition, the latency-sensitive application may be charged more than the reasonable price when the competition among the on-site spectrum requesters is fierce under the spectrum scarcity, giving rise to unfair trading.\par
\begin{figure}[htb!]
	\setlength{\abovecaptionskip}{-0cm}
	\setlength{\belowcaptionskip}{-0cm}
	\centering
	\includegraphics[width=0.50\textwidth]{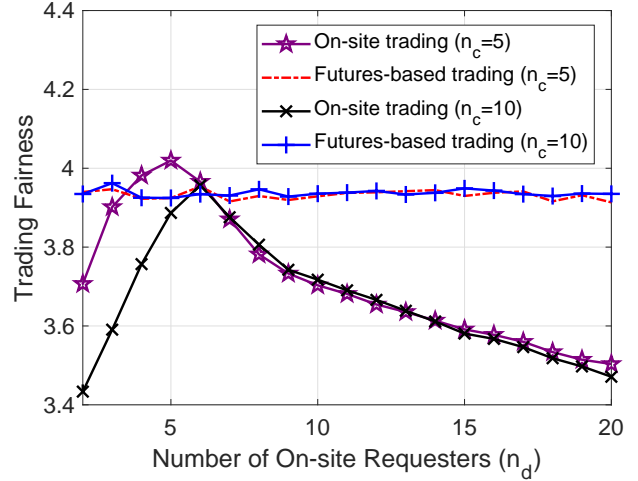}  
	\caption{Comparison of trading fairness of the on-site and futures-based trading scheme.}
	\label{fairness}
\end{figure}
The trading fairness of the proposed futures-based scheme remains almost constant and is higher than that of the on-site trading mechanism in most cases, as shown in Fig.~\ref{fairness}. This is because the spectrum amount for trading has been fixed through a forward contract. In contrast, the spectrum availability in the on-site trading is based on dynamic spectrum requester and local user numbers. The fairness of the on-site trading shows a change of trend from rise to decline with the increasing number of on-site requesters. The reason is that the available spectrum resource is greater than demand when the number of on-site requesters is small, leading to unfair trading to the spectrum owners. The spectrum is less than the overall demand and the spectrum access competition is intensified with the growth of on-site requesters. The trading is unfair to the spectrum requesters in this situation.\par
\section{Conclusion}
\label{ }
A futures-based resource trading scheme for real-time IoT networks is proposed to address the risk of trading failure and unfairness associated with the on-site trading. To achieve the futures-based trading, a resource requester and a resource owner sign a mutually beneficial forward contract, in which the risk of trading failure based on the statistics of available resource and demand is taken into account. With the proposed trading mechanism, a contract user obtains the resource immediately at the agreed price without the on-site negotiation delay. Simulation results confirm that the proposed futures-based resource trading outperforms the on-site resource trading in terms of trading success rate and fairness.\par

\end{document}